\documentclass[11pt,a4paper]{article}
\usepackage[utf8]{inputenc}
\usepackage[T1]{fontenc}
\usepackage{lmodern}
\usepackage{microtype}
\usepackage[margin=2.5cm]{geometry}
\usepackage{setspace}
\usepackage{parskip}
\usepackage{graphicx}
\usepackage{booktabs}
\usepackage{enumitem}
\usepackage{multicol}
\usepackage{natbib}
\usepackage{hyperref}
\hypersetup{colorlinks=true,linkcolor=blue,citecolor=blue,urlcolor=blue}
\usepackage{titling}
\usepackage{fancyhdr}
\usepackage{caption}
\usepackage{amsmath,amssymb}

\usepackage{aas_macros}
\usepackage{natbib}
\usepackage{ragged2e}
\usepackage{wrapfig}

\pretitle{\vspace*{-1cm}\begin{center}\LARGE\bfseries}
\posttitle{\par\end{center}\vskip 0.5em}
\preauthor{\begin{center}\large}
\postauthor{\end{center}}
\predate{\begin{center}\large}
\postdate{\end{center}}

\begin{document}

\thispagestyle{empty}

  \centering
  \vspace*{0.1cm}
  {\Huge\bfseries Expanding Horizons \\[6pt] \Large Transforming Astronomy in the 2040s \par}
    \vspace{0.2cm}

{\LARGE \textbf{Asteroseismology of white dwarfs in the 2040s}\par}
  \vspace{0.2cm}

\begin{tabular}{p{4.5cm}p{10cm}}
\textbf{Scientific Categories:} & Stellar evolution; Pulsating stars; Pre-white dwarfs and white dwarfs \\[0.4cm]

\textbf{Submitting Author:} &
Murat Uzundag \\
& Institute of Astronomy, KU Leuven, Celestijnenlaan 200D, B-3001 Leuven, Belgium \\
& \texttt{muratuzundag.astro@gmail.com}
\end{tabular}

\vspace{0.3cm}

\noindent
\textbf{Contributing authors:}  
Ingrid Pelisoli$^{1}$, Stephane Charpinet$^{2}$,  
Alejandro H. C\'orsico$^{3}$, Leandro G. Althaus$^{3}$, V. Van Grootel$^{4}$, Suzanna Randall$^{5}$,  Thomas Kupfer$^{6}$, Roberto Raddi$^{7}$\vspace{0.3cm}\\

{\raggedright\scriptsize
$^{1}$Department of Physics, University of Warwick, Gibbet Hill Road, Coventry, CV4 7AL, UK\\
$^{2}$Institut de Recherche en Astrophysique et Planétologie, CNRS, Université de Toulouse, France\\
$^{3}$Instituto de Astrofísica de La Plata, IALP (CCT La Plata), CONICET--UNLP, Argentina\\
$^{4}$Université de Liège, Belgium\\
$^{5}$European Southern Observatory, Germany\\
$^{6}$Hamburger Sternwarte, University of Hamburg, Germany\\
$^{7}$Universitat Politècnica de Catalunya, Departament de Física, Spain
\par}

\begin{abstract}

White dwarfs, the final evolutionary stage of the vast majority of stars, serve as critical tools for cosmochronology, studies of planetary system evolution, and laboratories for non-standard physics, including exotic cooling channels and weakly interacting particles, as well as crystallization processes. Beyond surface properties accessible via spectroscopy and model atmospheres, global pulsations exhibited by white dwarfs during various evolutionary phases provide a direct window into their deep interiors. Asteroseismology—the comparison of observed pulsation periods with theoretical models—enables us to infer internal chemical stratification, total mass, rotation profiles, and magnetic field strengths. Despite major advances from space missions providing uninterrupted, high-precision photometry, key challenges remain: many predicted pulsators remain quiet, while others oscillate outside theoretical instability strips, highlighting gaps in our understanding of mode excitation, diffusion, and convective mixing. Determining the masses of white dwarfs, particularly for massive and hydrogen-deficient stars, remains uncertain, with discrepancies between spectroscopic, asteroseismic, astrometric, and photometric methods. In the coming decades, large-scale surveys combining high-precision space-based photometry with coordinated ground-based spectroscopic follow-up will dramatically increase both the number and quality of pulsating white dwarf observations. These comprehensive datasets will enable population-level asteroseismic studies, providing robust constraints on internal rotation, core composition, and crystallization, and ultimately refining models of stellar evolution. A coordinated effort in the 2040s will fully exploit these opportunities, systematically probing the internal physics of pulsating white dwarfs across stellar populations.

\end{abstract}

\newpage

\normalfont\normalsize
\justifying


\section{Introduction and Background}
\label{sec:intro}

White dwarfs (WDs) represent the final evolutionary stage of stars with initial masses below $\sim7$–$11\,M_{\odot}$, encompassing over 95\% of Milky Way stars, including the Sun \citep{2001PASP..113..409F,2010A&ARv..18..471A}. As such, they are key laboratories for studying stellar evolution, planetary system remnants \citep{2019Natur.576...61G}, crystallization \citep{2019Natur.565..202T}, and non-standard physics, including the properties of exotic particles like axions, and the secular variation of the fundamental constants of physics \citep{2022FrASS...9....6I}. A subset of WDs exhibit photometric variability due to pulsations, binarity, or transiting planetary debris \citep[e.g.,][]{2021ApJ...912..125G}. Among these, pulsations provide a unique window into stellar interiors through asteroseismology —the comparison of observed pulsation periods with theoretical periods— allowing precise determination of internal structure, chemical stratification, total mass, rotation profiles, and magnetic fields \citep[e.g.,][]{2019A&ARv..27....7C}.

WD pulsations are global, non-radial $g$-modes with harmonic degrees $\ell=1,2$, pulsation periods $\sim100$–7000\,s, and amplitudes typically below 0.4\,mag \citep{2008ARA&A..46..157W}. Excitation mechanisms vary with composition and evolutionary stage, arising from partial ionization zones ($\kappa-\gamma$ mechanism) \citep{1981A&A...102..375D, 1982ApJ...252L..65W} or convection in outer layers \citep{1991MNRAS.251..673B}. Pulsations are highly sensitive to internal chemical profiles, enabling inference of key parameters such as $T_{\rm eff}$, $\log g$, and $M_\star$, as well as core composition, crystallization, and envelope layer thickness through mode trapping \citep[e.g.,][]{2023MNRAS.526.2846U}. Frequency splitting further constrains rotation and magnetic fields, and pulsating WDs can even probe fundamental physics, including axions, neutrinos, and general relativity \citep{2023MNRAS.524.5929C}.

\begin{figure}[t]
  \centering
  \includegraphics[width=0.75\columnwidth]{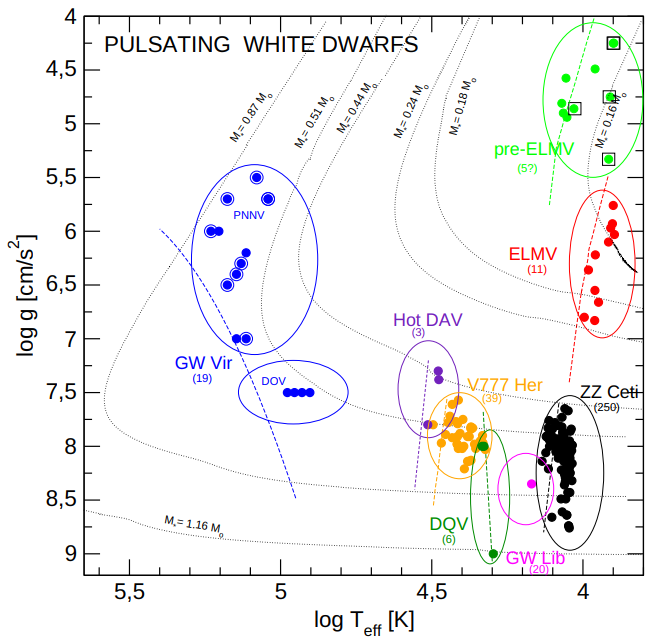}
  \caption{
  Location of confirmed and candidate pulsating white dwarfs and pre--white-dwarf stars in the $\log T_{\rm eff}$--$\log g$ diagram, adapted from \citet{2019A&ARv..27....7C}. Symbols denote different variability classes, including DAV, DBV, GW~Vir, ELMV, and pre-ELMV stars; GW~Vir planetary nebula nucleus variables are highlighted. Representative evolutionary tracks for low-mass He-core, H-deficient post-VLTP, and ultramassive H-rich white dwarfs are shown for reference, together with theoretical blue edges of the instability domains (dashed lines).
  }
  \label{fig:wd_pulsators}
\end{figure}

The majority of observed WDs ($\sim$70–80\%) are H-rich (DA), hosting the most common pulsators: ZZ Ceti (DAV) stars. These stars pulsate in a narrow instability strip at $T_{\rm eff}\sim10\,500$–$13\,000$\,K and $\log g\sim7.5$–9.2. Space-based high-precision photometry has revolutionized their study, revealing rotation periods, outbursts near the red edge, and mode-line width dichotomies \citep{2017ApJS..232...23H}. Observations have expanded the DAV sample to over 500 stars, including ELMVs (He-core extremely low-mass variables) produced by binary evolution \citep{2025A&A...699A.280A}, and the hotter “hot DAVs” at $T_{\rm eff}\sim30\,000$\,K \citep{2020MNRAS.497L..24R}. Theoretical predictions also suggest a “warm” DA pulsator population ($T_{\rm eff}\sim19\,000$\,K) yet to be observed \citep{2020A&A...633A..20A}. Figure \ref{fig:wd_pulsators} summarizes the location of the various classes of pulsating white dwarfs and pre–white-dwarf stars in the $\log T_{\rm eff}$–$\log g$ plane, together with representative evolutionary tracks and theoretical instability boundaries.

H-deficient WDs ($\sim$20–30\%) span spectral types PG~1159, DO, DB, and DQ, and exhibit more diverse and less understood pulsational behavior. Key classes include GW Vir stars (hot, luminous pre-WDs descended from PG~1159 progenitors) and DBVs (V777 Her stars, $T_{\rm eff}\sim25\,000$\,K). Their evolutionary pathways—single-star channels versus WD mergers—remain uncertain, and theoretical instability strips often do not match observations \citep{2019A&ARv..27....7C,2022ApJ...927..158V}. Space-based photometry has enabled the discovery of new pulsators, detailed period analyses, rotational splitting, and comparison of asteroseismic distances with Gaia measurements \citep{2021A&A...645A.117C,2022A&A...668A.161C,2021A&A...655A..27U,2022MNRAS.513.2285U}. These studies continue to refine our understanding of internal structure, rotation, and evolutionary history for these rare objects.

\section{Open Science Questions in the 2040s}
\label{sec:openquestions}

By the 2040s, large-scale surveys combining space-based photometry with ground-based spectroscopy will enable systematic progress on several long-standing questions in the physics of pulsating white dwarfs:

\textbf{Mode scarcity in DAVs and DBVs and incomplete rotational multiplets:}  
Observed pulsators show far fewer modes than predicted, likely due to very low amplitudes. Detecting additional modes is essential for reliable asteroseismic modeling and for understanding mode excitation, selection, and amplitude limitation, requiring high-precision photometry with large telescopes. Moreover, many stars exhibit rotationally split modes with missing components. Clarifying why some modes form complete multiplets while others do not is key to constraining internal rotation and angular momentum transport.

\textbf{Impure instability strips:}  
The presence of non-variable stars within theoretical instability strips, particularly among ultramassive DAVs, may reflect undetected low-amplitude pulsations or uncertainties in $T_{\rm eff}$ and $\log g$. Improved sensitivity and atmospheric parameter determinations are needed.

\textbf{Hydrogen and helium envelope masses:}  
The thickness of H envelopes in DAVs and He envelopes in DBVs remains poorly constrained. Ensemble asteroseismology, based on large samples with many detected modes, will enable statistical constraints on envelope properties.

\textbf{Chemical interfaces and mode trapping:}  
Chemical stratification in DAV, DBV, and GW~Vir stars produces mode trapping signatures in period spacing patterns. Recovering these diagnostics requires high-quality detections of numerous pulsation modes.

\textbf{Outbursting white dwarfs:}  
Outbursts observed near the red edge of the DAV instability strip, and possibly in DBVs, remain poorly understood. Expanded samples will allow population-level studies of their incidence and physical origin.

\textbf{Ultramassive pulsating DA white dwarfs:}  
WDs with $M_\star \gtrsim 1.05\,M_\odot$ probe crystallization, core composition (CO versus ONe), super-AGB evolution, and merger remnants. Dense mode spectra will test crystallization theory, envelope structure, and formation channels.

\textbf{Mass discrepancies across methods:}  
Systematic disagreements between spectroscopic, photometric, astrometric, and seismological mass estimates persist across DAV, DBV, and GW~Vir stars. These discrepancies reflect uncertainties in atmospheric modeling, cooling and mass--radius relations, and asteroseismic degeneracies arising from core--envelope symmetry and limited mode sampling. Resolving them requires large, homogeneous samples and improved evolutionary and pulsation models.

\section{Technology and Data Handling Requirements}
\label{sec:tech}

Low- to mid-resolution spectroscopy (R $\approx$ 2{,}000--10{,}000) across 350--900\,nm is essential for deriving accurate atmospheric parameters ($T_{\rm eff}$, $\log g$, and surface composition) for large samples of faint pulsating white dwarfs and pre--white-dwarf stars. These measurements provide the fundamental inputs for asteroseismic modelling by tightly constraining stellar mass, envelope structure, and the location of chemical transition zones. 

High-resolution spectroscopy (R $\gtrsim$ 20{,}000--40{,}000) obtained with 4--10\,m-class telescopes further enables the detection of weak metal lines, the characterization of diffusion processes, and detailed line-profile variability studies associated with pulsation, offering complementary constraints on internal chemical stratification and envelope physics. Access to large-aperture facilities is also critical for achieving sufficient signal-to-noise ratios for the faintest targets, including ultra-massive pulsating white dwarfs ($G \approx 17$--22\,mag). 

In parallel, large-domain time-series photometric surveys providing long-baseline, multi-colour observations will play a transformative role in white dwarf asteroseismology. Multi-band photometry enables mode identification through amplitude and phase differences, while repeated monitoring over several years allows the detection of secular changes in pulsation periods. This is particularly relevant for rapidly evolving objects such as GW~Vir stars, which are expected to exhibit measurable period drifts driven by cooling and contraction on timescales of only a few years. Long-term monitoring of these stars with repeated observations on a roughly annual timescale enables direct observational tests of evolutionary models through measurements of $\dot{P}$, as demonstrated for GW~Vir pulsators.

\begin{multicols}{2}
\tiny	
\setlength{\bibsep}{1pt}

\bibliographystyle{mnras}
\bibliography{refs}

\end{multicols}

\end{document}